\pdfoutput=1   % for arXiv

\documentclass[usenatbib,usegraphicx]{mn2e}
\usepackage{amssymb,bm,color,multirow,charter}
\newcommand{\e}[1]{\times 10^{#1}}                        % x 10^...
\renewcommand{\vec}[1]{\bm{\mathit #1}}                          % vector
\newcommand{\fig}[1]{Fig. \ref{fig:#1}}
\newcommand{\sect}[1]{\S\ref{sec:#1}}
\newcommand{\tab}[1]{Table \ref{tab:#1}}

\newcommand{\fignar}[3]{
   \begin{figure}
   \centering
   \includegraphics{#1}
   \caption{#3}
   \label{fig:#2}
   \end{figure}
}

\newcommand{\figwide}[3]{
   \begin{figure*}
   \centering
   \includegraphics{#1}
   \caption{#3}
   \label{fig:#2}
   \end{figure*}
}

\title[QPOs from Unstable Accretion: 3D MHD Simulations]{Possible Quasi-Periodic Oscillations from Unstable Accretion: 3D MHD Simulations}
\author[A. K. Kulkarni \& M. M. Romanova]
{A. K. Kulkarni\thanks{E-mail: akshay@astro.cornell.edu},
M. M. Romanova\thanks{E-mail: romanova@astro.cornell.edu}\\
Dept. of Astronomy, Cornell University, Ithaca, NY 14853}

\begin{document}
\maketitle

\begin{abstract}
We investigate the photometric variability of magnetized stars, particularly neutron stars, accreting through a magnetic Rayleigh-Taylor-type instability at the disk-magnetosphere interface, and compare it with the variability during stable accretion, with the goal of looking for possible quasi-periodic oscillations. The lightcurves during stable accretion show periodicity at the star's frequency and sometimes twice that, due to the presence of two funnel streams that produce antipodal hotspots near the magnetic poles. On the other hand, lightcurves during unstable accretion through tongues penetrating the magnetosphere are more chaotic due to the stochastic behaviour of the tongues, and produce noisier power spectra. However, the power spectra do show some signs of quasi-periodic variability. Most importantly, the rotation frequency of the tongues and the resulting hotspots is close to the inner-disk orbital frequency, except in the most strongly unstable cases. There is therefore a high probability of observing QPOs at that frequency in longer simulations. In addition, the lightcurves in the unstable regime show periodicity at the star's rotation frequency in many of the cases investigated here, again except in the most strongly unstable cases which lack funnel flows and the resulting antipodal hotspots. The noisier power spectra result in the fractional rms amplitudes of the Fourier peaks being smaller.

We also study in detail the effect of the misalignment angle between the rotation and magnetic axes of the star on the variability, and find that at misalignment angles $\gtrsim 25^\circ$, the star's period always appears in the lightcurves.

% We investigate the photometric variability of \referee{magnetized stars, particularly} neutron stars, accreting through a magnetic Rayleigh-Taylor-type instability at the disk-magnetosphere interface, and compare it with the variability during stable accretion, with the goal of looking for possible quasi-periodic oscillations. \referee{The lightcurves during unstable accretion generally show periodicity at the star's rotation frequency, and also signs of quasi-periodic variability, except in the most strongly unstable cases.} The power spectra are noisier than during stable accretion, with the result that the fractional rms amplitudes of the Fourier peaks are smaller.
% 
% We also study in detail the effect of the misalignment angle between the rotation and magnetic axes of the star on the variability, and find that at misalignment angles $\gtrsim 25^\circ$, the star's period always appears in the lightcurves.
\end{abstract}

\begin{keywords}
accretion, accretion discs; instabilities; MHD; stars:neutron; stars: oscillations; stars: magnetic fields
\end{keywords}

%=========================================================

\section{Introduction}
Accretion to magnetized stars occurs in different systems, including classical T Tauri stars (CTTSs), which are the progenitors of solar-type stars \citep[e.g.,][]{BouvierEtAl07}, in magnetized cataclysmic variables, which are accreting white dwarfs \citep[e.g.,][]{Warner95}, and in millisecond pulsars, which are weakly magnetized accreting neutron stars \citep[e.g.,][]{vanderKlis04}. The lightcurves of these stars are expected to show definite periodicity corresponding to the star's rotation period, but in addition to this they often show interesting features like quasi-periodic oscillations (QPOs), or lack any periodicity altogether. The reasons for this behaviour are not yet perfectly clear. The lack of periodicity in many CTTSs was thought to be due to the lack of a dynamically important magnetic field, but recent observations have shown that that is not always the case \citep[e.g.,][]{DonatiEtAl07}. QPO models in neutron star systems usually attempt to look towards characteristic frequencies in the disk, and their beats with the stellar rotation frequency, for explanations \citep{vanderKlis04}. The variability characteristics of dwarf novae, which are a sub-class of cataclysmic variables, show close similarity to those of neutron stars \citep{WarnerWoudtPretorius03}. These observations hint at a common explanation for the variability features in these systems, perhaps based on the properties of the accretion flow itself. This leads us to investigate the effect of the accretion flow on the variability.

Accretion to magnetized stars can be in one of two regimes: (i) the stable regime, in which the accretion disk is stopped by the star's magnetic field, and the accreting matter flows around the magnetosphere, forming two funnels that deposit matter near the star's magnetic poles \citep{GhoshLamb79, Konigl91, KoldobaEtAl02, RomanovaEtAl03, RomanovaEtAl04}, or (ii) the unstable regime, in which Rayleigh-Taylor-type instabilities occur at the disk-magnetosphere interface \citep{AronsLea76, ElsnerLamb77, WangRobertson85}. The instabilities produce tongues of matter that penetrate the magnetosphere and deposit matter much closer to the star's equator \citep[hereafter KR08]{RomanovaKulkarniLovelace08, KulkarniRomanova08}. The tongues rotate around the star with an angular velocity close to the inner-disk orbital frequency, and their number is of the order of a few. In KR08 it was noticed that the lightcurves during unstable accretion can lack the clear periodicity seen in the lightcurves during stable accretion \citep[see, e.g.,][hereafter KR05]{PoutanenGierlinski03, KulkarniRomanova05}. Here we investigate this behaviour in more detail in accreting magnetized stars, particularly millisecond pulsars. Our goal is to look for signs of QPOs in the unstable regime.

KR08 also found that the instability is suppressed at high misalignment angles between the star's rotation and magnetic axes. Here we perform a detailed investigation of the boundary between stable and unstable regimes as a function of the misalignment angle and stellar rotation rate, focusing on whether the lightcurves show definite periodicity.

We start out with a summary of our previous 3D MHD simulation results in \sect{sum-prev} and a description of our variability model in \sect{model-variab}. This is followed in \sect{variab} by a comparison of variability features during stable and unstable accretion, including a discussion of the instability on the accretion rate, misalignment angle and stellar rotation rate. We also discuss signs of QPO-like behaviour. We end in \sect{conc} with our conclusions and a discussion of the implications of this work for observations of accreting millisecond pulsars.

%=========================================================

\section{Summary of 3D MHD Simulation Results}
\label{sec:sum-prev}
For context, we summarize the model and results of our 3D MHD simulations described in KR08.

%---------------------------------------------------------

\subsection{Model}
\label{sec:model-3d}
The model we use is the same as in our earlier 3D MHD simulations (Koldoba et al. 2002; Romanova et al. 2003, 2004). The star has a dipole magnetic field, the axis of which makes an angle $\Theta$ with the star's rotation axis. The rotation axes of the star and the accretion disk are aligned. There is a low-density corona around the star and the disk which also rotates about the same axis. To model stationary accretion, the disk is chosen to initially be in a quasi-equilibrium state, where the gravitational, centrifugal and pressure gradient forces are in balance (Romanova et al. 2002). Simulations were done for both relativistic and non-relativistic stars. General relativistic effects, which are important for neutron stars, are modelled using the Paczy\'nski-Wiita potential $\Phi(r) = GM/(r-r_g)$ (Paczy\'nski \& Wiita 1980), where $M$ is the mass of the star and $r_g \equiv 2GM/c^2$ is its Schwarzschild radius. This potential reproduces some important features of the Schwarzschild spacetime, such as the positions of the innermost stable and marginally bound circular orbits. Viscosity is modelled using the $\alpha$-model (Shakura \& Sunyaev 1973; Novikov \& Thorne 1973), and controls the accretion rate through the disk. To model accretion, the ideal MHD equations are solved numerically in three dimensions, using a Godunov-type numerical code, written in a ``cubed-sphere'' coordinate system rotating with the star (Koldoba et al. 2002; Romanova et al. 2003). The boundary conditions at the star's surface amount to assuming that the infalling matter passes through the surface of the star. So the dynamics of the matter after it falls on the star is ignored. The inward motion of the accretion disk is found to be stopped by the star's magnetosphere at the Alfv\'en or magnetospheric radius, where the magnetic and matter energy densities become equal. The subsequent evolution depends on the parameters of the model.

%---------------------------------------------------------

\subsection{Reference Values}
\label{sec:refval}
The simulations are done using dimensionless variables. For every physical quantity $q$, the dimensionless value is defined as $q' = q/q_0$, where $q_0$ is the reference value for $q$. Because of the use of dimensionless variables, the results are applicable to a wide range of objects and physical conditions, each with its own set of reference values, provided the magnetospheric radius $r_m$ is upto a few times the radius of the star $R$; $r_m$ is determined by the balance of the magnetospheric and matter pressures, so that the modified plasma parameter at the disk-magnetosphere boundary $\beta=(p+\rho v^2)/(B^2/{8\pi})\approx 1$. To apply our simulation results to a particular situation, we have the freedom to choose three parameters, and all the reference values are calculated from those. We choose the mass, radius and equatorial surface magnetic field of the star as the three independent parameters. Appendix \ref{app:refval} shows how the reference values are determined, and lists the reference values for three classes of central objects --- classical T Tauri stars, white dwarfs and neutron stars.

Subsequently, we drop the primes on the dimensionless variables and show dimensionless values everywhere unless otherwise stated.

%---------------------------------------------------------

\subsection{Results}
We found that accretion to magnetized stars with magnetospheres a few times the stellar radius in size can be either stable or unstable. In the stable regime, the accretion disk is stopped by the star's magnetic field, and the accreting matter flows around the magnetosphere, forming two funnels that deposit matter near the star's magnetic poles. This produces two antipodal hotspots, and the lightcurves show periodicity at either the star's frequency or twice that, depending on the viewing geometry. In the unstable regime, Rayleigh-Taylor-type instabilities occur at the disk-magnetosphere interface, producing tongues of matter that penetrate the magnetosphere and deposit matter much closer to the star's equator. The tongues rotate around the star with an angular velocity close to the inner-disk orbital frequency, and their number is of the order of a few. The lightcurves often lack the clear periodicity seen during stable accretion. However, the fact that the tongues rotate at approximately the inner-disk orbital frequency leads to the hope of observing some periodicity or quasi-periodicity at that frequency.

The instability is associated with high accretion rates, and coexists with funnel flows for a relatively broad range of accretion rates. It is expected to occur in most accreting systems for typical values of mass, radius, surface magnetic fields and accretion rates. We quote here from KR08 the empirically obtained critical accretion rate, measured at the stellar surface, separating the stable and unstable accretion regimes (see KR08 for a more detailed discussion of instability criteria):

\begin{equation}
\label{eq:mcrit}
\dot M_{crit} = \dot M_{crit,0} \left(\frac{B}{B_0} \right)^2 \left(\frac{R}{R_0} \right)^{5/2} \left(\frac{M}{M_0} \right)^{-1/2},
\end{equation}
where $B$, $R$ and $M$ are the surface magnetic field, radius and mass of the star respectively (in dimensional units), and the subscript `0' denotes the reference values shown in Table \ref{tab:refval}. The characteristic critical accretion rate $\dot M_{crit,0}$ depends on the magnetospheric radius, which is set by the dimensionless value of the stellar magnetic dipole moment $\mu$ and is (4--5)$R$ for $\mu=2$ and (2--3)$R$ for $\mu=0.5$. The corresponding values of $\dot M_{crit,0}$ for classical T Tauri stars, white dwarfs and neutron stars are given in Table \ref{tab:mdotcrit}.

\begin{table}
\begin{centering}
\begin{tabular}{llll}
%________________________________________________________________________
\hline
                   & CTTSs           & White dwarfs    & Neutron stars \\
%________________________________________________________________________
\hline
$r_m=$ (2--3)$R$   &   $1.7\e{-7}$   &   $1.1\e{-7}$   &   $1.8\e{-8}$ \\
$r_m=$ (4--5)$R$   &   $2.1\e{-8}$   &   $1.4\e{-8}$   &   $2.2\e{-9}$ \\
%________________________________________________________________________
\hline
\end{tabular}
\end{centering}
\caption{Characteristic critical accretion rates $\dot M_{crit,0}$ ($M_\odot \mbox{ yr}^{-1}$) for various types of accreting systems, for two values of the magnetospheric radius $r_m$.}
\label{tab:mdotcrit}
\end{table}

%=========================================================

\section{Variability Model}
\label{sec:model-variab}

We calculate lightcurves from the accretion hotspots on the stellar surface, taking into account the general relativistic effects of light bending and gravitational redshift, and the Doppler effect, which are important for neutron stars. Light travel time effects are negligible and are not taken into account (\citealt{PoutanenGierlinski03}; KR05). To calculate the emission from the hotspots, we integrate the emitted flux over the star's surface, using the method described by KR05. The observed flux is then given by
\begin{equation}
F = \left(1-\frac{r_g}{R}\right)^2 \int_{\cos\lambda>0} dS' \frac{I_E'(\vec R',\lambda)\cos\lambda}{\gamma^5(1-\beta\mu)^5}.
\end{equation}
Here, primed and unprimed quantities are measured in the star's and observer's rest frames respectively. $R$ and $r_g$ are the star's radius and Schwarzschild radius respectively. For a point on the star's surface rotating with velocity $v$, $\beta\equiv v/c$ and $\gamma\equiv (1-\beta^2)^{-1/2}$. $I_E$ is the frequency-integrated specific intensity of the emitted radiation. Finally, $\lambda$ is the angle between the local radial direction and the direction in which a light ray needs to be emitted so as to reach the observer, taking light bending into account.

We now assume that the entire kinetic and thermal energy of the matter falling onto the star's surface is converted into blackbody radiation, which is emitted isotropically. Then the flux emitted from the star's surface is
\begin{equation}
F_E(\vec R',\lambda) = \pi I_E(\vec R',\lambda),
\end{equation}
giving,
\begin{equation}
F = \frac{1}{\pi}\left(1-\frac{r_g}{R}\right)^2 \int_{\cos\lambda>0} dS' \frac{F_E'(\vec R',\lambda)\cos\lambda}{\gamma^5(1-\beta\mu)^5}.
\end{equation}

\fignar{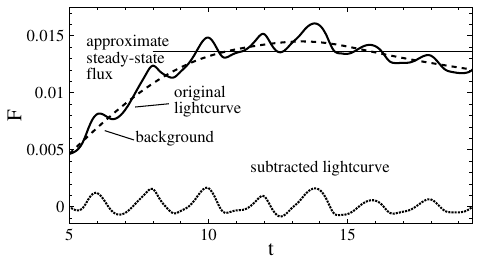}{bksub-demo}{Hotspot lightcurve from one of our simulations (solid line), the slowly varying background flux (dashed line), the difference (dotted line), and the approximate steady-state flux (solid horizontal line).}

\fignar{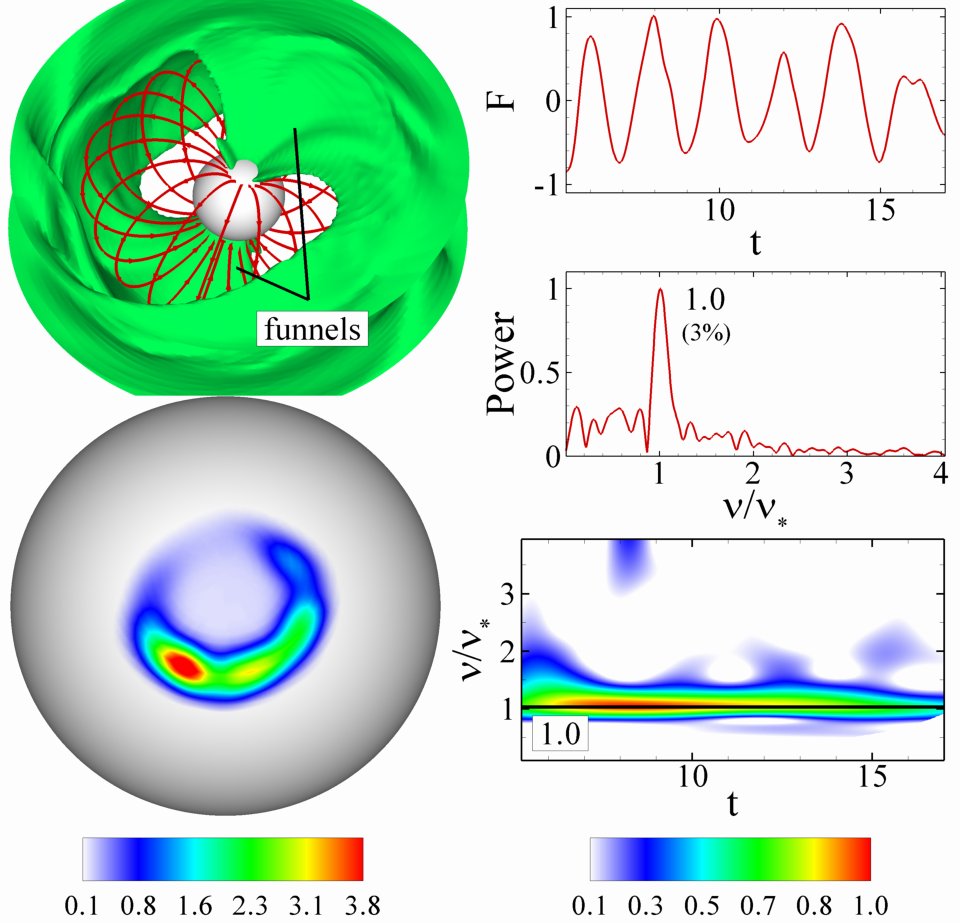}{stable1}{Stable accretion case which has the following parameters: viscosity parameter $\alpha=0.02$, misalignment angle $\Theta=15^\circ$, stellar rotation period $P=2.0$, stellar magnetic moment $\mu=1$. The observer inclination angle is $i=15^\circ$. {\it Top left:} Constant plasma density surface showing matter flow around star, and magnetospheric magnetic field lines. {\it Bottom left:} Northern hotspot, as seen from above the magnetic pole, showing the emitted flux. {\it Right column:} Hotspot lightcurve (normalized) and its Fourier (normalized) and wavelet spectra. The colours in the hotspot and wavelet plots range from white (low) through blue to red (high). The numbers inside the Fourier and wavelet plots are values of $\nu/\nu_*$, and the percentages are fractional rms amplitudes.}

\figwide{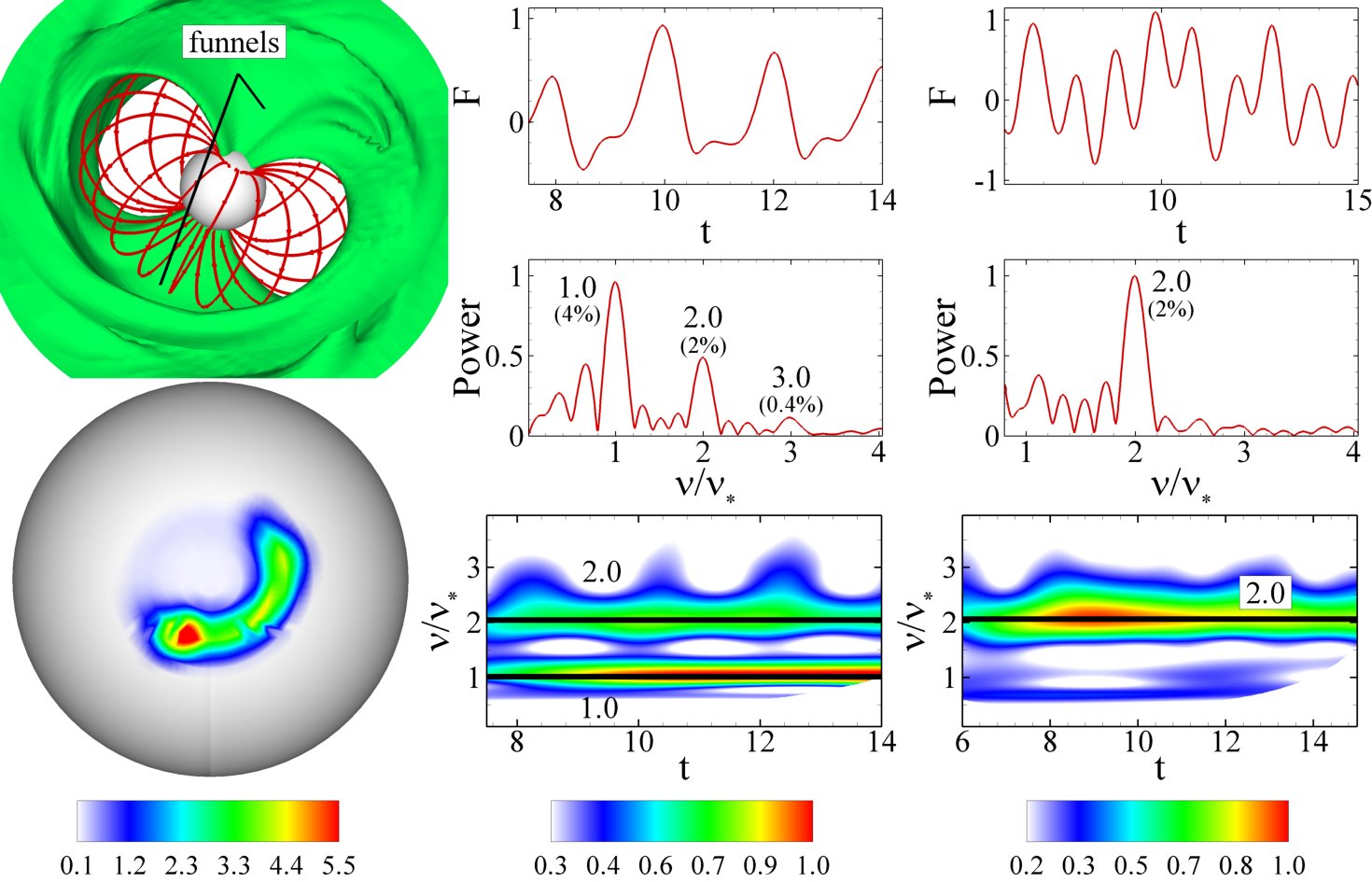}{stable12}{Similar plot to \fig{stable1} for a stable accretion case which has the following parameters: $\alpha=0.02$, $\Theta=30^\circ$, $P=2.0$, $\mu=1$. The observer inclination angle is $i=45^\circ$ and $i=90^\circ$ in the second and third columns respectively.}

Before performing the variability analysis, one more thing needs to be done. \fig{bksub-demo} shows the hotspot lightcurve from one of our simulations. Periodic variability is clearly seen superposed on a slowly varying background. The variation of this background depends on the long-term behaviour of the disk in our simulations. However, this is not of interest for the short-term variability considered in the present work. We therefore smooth the lightcurve to obtain the background and subtract it from the lightcurve before analyzing the variability. This makes the short-term variability much clearer. In all subsequent figures, we show only the subtracted lightcurves. We estimate the pulsed fraction of the emission using an approximate value for the background lightcurve flux in the steady state.

We show dimensionless values for all quantities in this work. To convert them into physical units, they need to be multiplied by the reference values listed in \tab{refval}. In particular, the values of time are in units of the Keplerian period $P_0$ at $r=R_0=3R$, and not of the stellar rotation period, since we consider stars with different rotation rates. The observer inclination angle varies from $0^\circ$ at the north rotational pole of the star to $90^\circ$ in the equatorial plane.

One final note: the lightcurves presented in this work are calculated for strongly relativistic systems like neutron stars, and are therefore not applicable to other types of accreting magnetized stars like classical T Tauri stars or protostars. However, the analysis of hotspot motion on the stellar surface in section \sect{spot-omega} uses the direct results of the 3D MHD simulations. These simulations, as mentioned earlier, approximate general relativistic effects using the Paczy\'nski-Wiita potential, but simulations without taking these effects into account show that the instability behaves in a very similar manner. We therefore expect the analysis of hotspot motion to be applicable to these two types of accreting systems as well.

%=========================================================

\section{Variability in the Stable and Unstable Regimes}
\label{sec:variab}

\subsection{Stable Accretion}
\label{sec:variab-stable}

\fignar{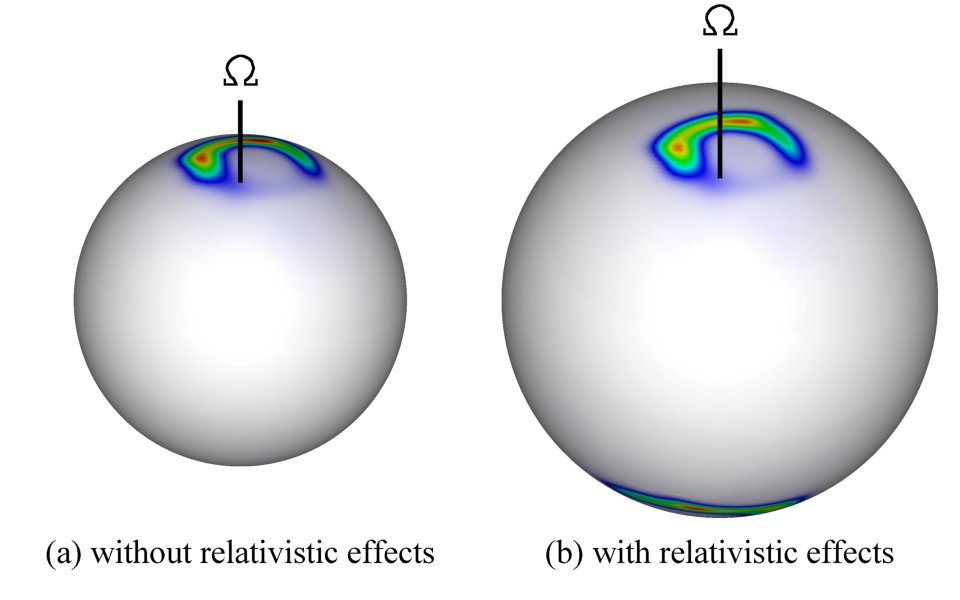}{img-gr}{Comparison of the visible portion of the neutron star with and without relativistic effects.}

In the stable regime, the funnels produce two antipodal hotspots near the star's magnetic poles. The funnels are steady, and therefore the hotspots are almost fixed on the star's surface (\citealt{RomanovaEtAl04}; KR05). Depending on the misalignment and inclination angles, either one or both antipodal hotspots can be seen by the observer during each rotation of the star. Correspondingly, the power spectrum shows peaks at either $\nu_*$, $2\nu_*$ or both. \fig{stable1} shows the lightcurve and its Fourier and wavelet\footnote{Since the wavelet transform uses a window of width $\Delta t$ centred at time $t$ to calculate the time-dependent frequency spectrum of the lightcurve from time $t_1$ to time $t_2$, it is necessary to have $t_1+\Delta t/2 \lesssim t \lesssim t_2 - \Delta t/2$. This results in a portion of the wavelet plot being cut off. The width $\Delta t$ is inversely related to the frequency, so lower frequencies are more strongly affected by this restriction.} spectra for one of our stable accretion cases, which has the following parameters: misalignment angle $\Theta=15^\circ$, viscosity parameter $\alpha=0.02$, stellar rotation period $P=2.0$ and stellar magnetic dipole moment $\mu=1$. The observer inclination angle is $i=15^\circ$. In this case, as the star rotates, only the northern hotspot is visible to the observer. So the hotspot lightcurve is modulated at the star's rotation frequency. As the wavelet shows, there is no drift in the pulsation frequency. The fractional rms amplitudes are small because the hotspots are near the rotational axis, and also due to general relativistic effects (\citealt{PoutanenGierlinski03}; KR05).

\fig{stable12} shows another stable accretion case, with $\Theta=30^\circ$ and $i=45^\circ$ (second column). For this accretion geometry, the frequencies $\nu_*$ and $2\nu_*$ are both visible. If the inclination angle $i=90^\circ$, only $2\nu_*$ is seen (\fig{stable12}, third column).

\fignar{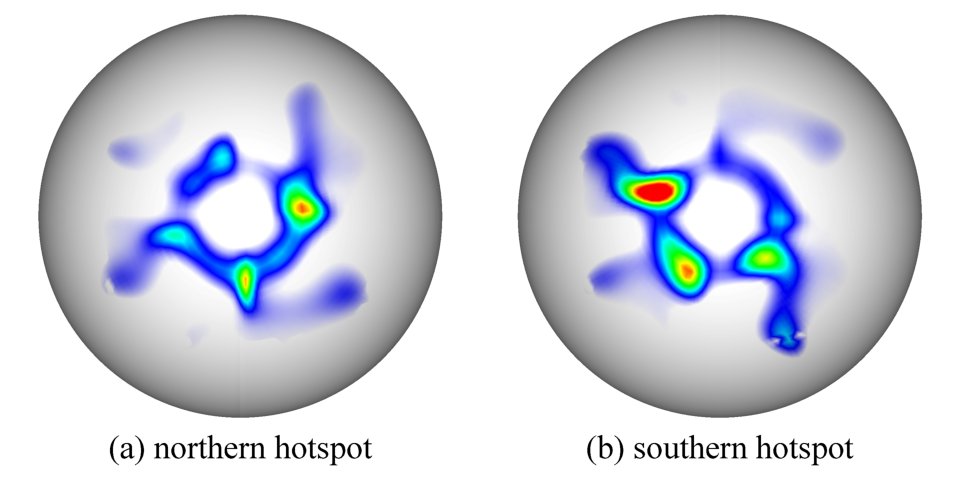}{north-south-spots}{Hotspots around the north and south magnetic poles for the unstable case in \fig{unstable} (discussed in \sect{variab-unstable}).}

One thing that needs to be noted, though, is that even in this relatively simple accretion geometry, it is not always straightforward to guess if the power spectrum will have peaks at $\nu_*$ or $2\nu_*$. Firstly, the hot spots are not exactly at the magnetic poles; rather, they are banana-shaped and surround the pole. Secondly, in neutron stars, due to gravitational bending of light, more than half the star's surface is visible, causing the hotspot on the far side of the star to be visible either for some or all stellar rotation phases (\fig{img-gr}; see also \citealt{Beloborodov02, PoutanenGierlinski03}; KR05). Thirdly, the two antipodal hotspots are not always identical (\fig{north-south-spots}; identical spots would have appeared as mirror images of each other here), in which case the fundamental frequency is $\nu_*$ and not $2\nu_*$, even at high observer inclination angles.

%---------------------------------------------------------

\subsection{Unstable Accretion}
\label{sec:variab-unstable}

\fignar{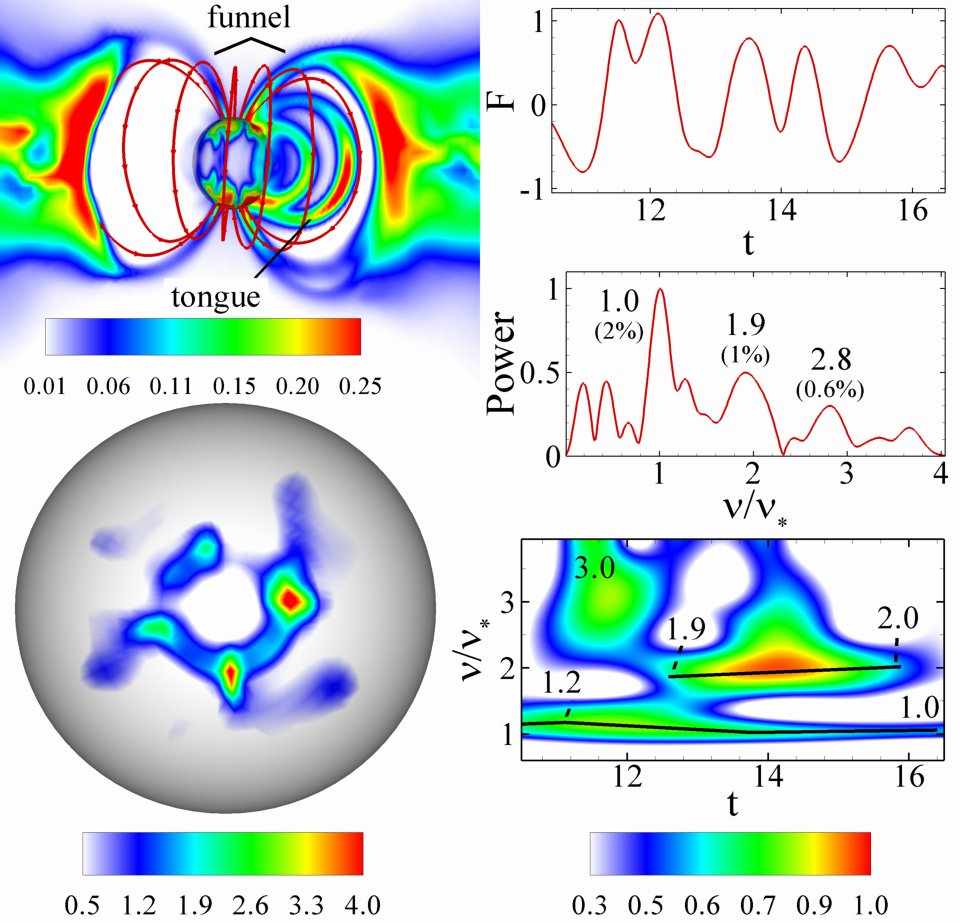}{unstable}{Similar plot to \fig{stable1} for an unstable case, which has the following parameters: $\alpha=0.02$, $\Theta=5^\circ$, $P=2.0$, $\mu=1$. The observer inclination angle is $i=45^\circ$. The matter flow in the top left panel is shown by density contours on two vertical slices passing through the origin.}

As we move into the unstable regime, we start seeing accretion through both funnels and tongues. The hotspots become more chaotic (\fig{unstable}). The funnel component of the hotspots still produces a peak at the star's rotation frequency in the lightcurve, but it is not as steady as in the stable case. The power spectrum shows additional peaks produced by the tongue component of the hotspots. These frequencies appear only sporadically, though, as the wavelet shows. This is because of the stochastic nature of the tongues.

Deeper in the unstable regime (\fig{unstable-strongly}), the hotspots are created entirely by tongues. Therefore, their shape, intensity, number and position changes on the inner-disk dynamical timescale. The star's rotation frequency is absent from the power spectrum, and the Fourier and wavelet spectra are very chaotic. An important exception to this are cases with small magnetospheres, which show clear QPO features \citep{RomanovaKulkarni08}.

\fignar{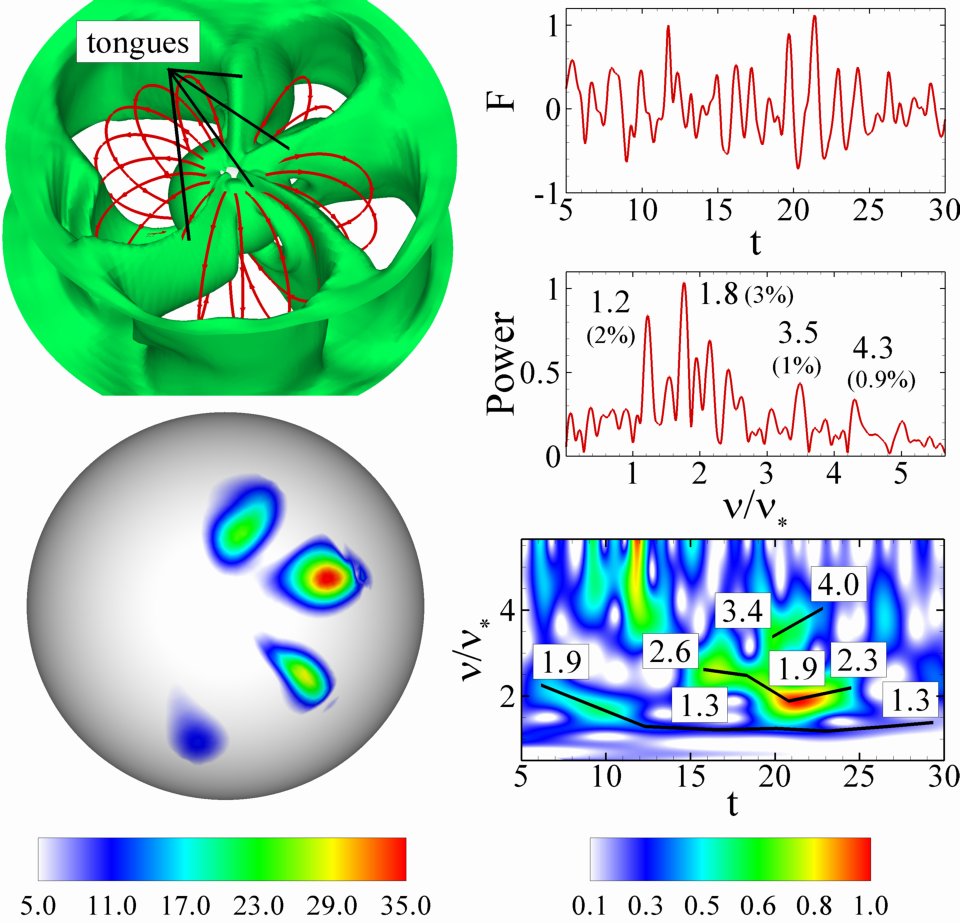}{unstable-strongly}{Similar plot to \fig{stable1} for a strongly unstable accretion case, which has the following parameters: $\alpha=0.2$, $\Theta=5^\circ$, $P=2.8$, $\mu=2$. The observer inclination angle is $i=45^\circ$.}

%---------------------------------------------------------

\figwide{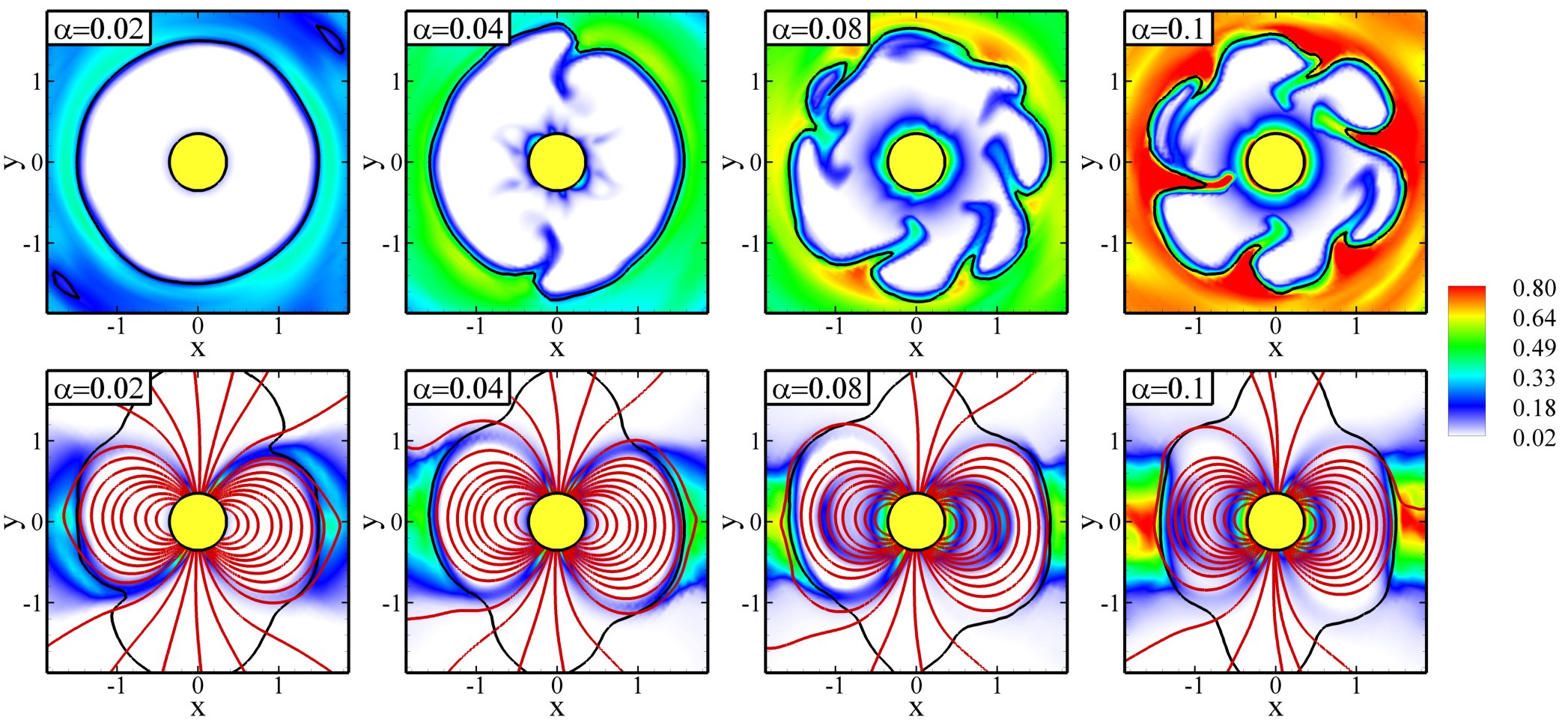}{xy-xz}{Equatorial (top row) and vertical (bottom row) slices showing the density distribution for various values of $\alpha$. The colours range from white (low) through blue to red (high). The black line denotes where the matter and magnetic energy densities are roughly equal.}

\figwide{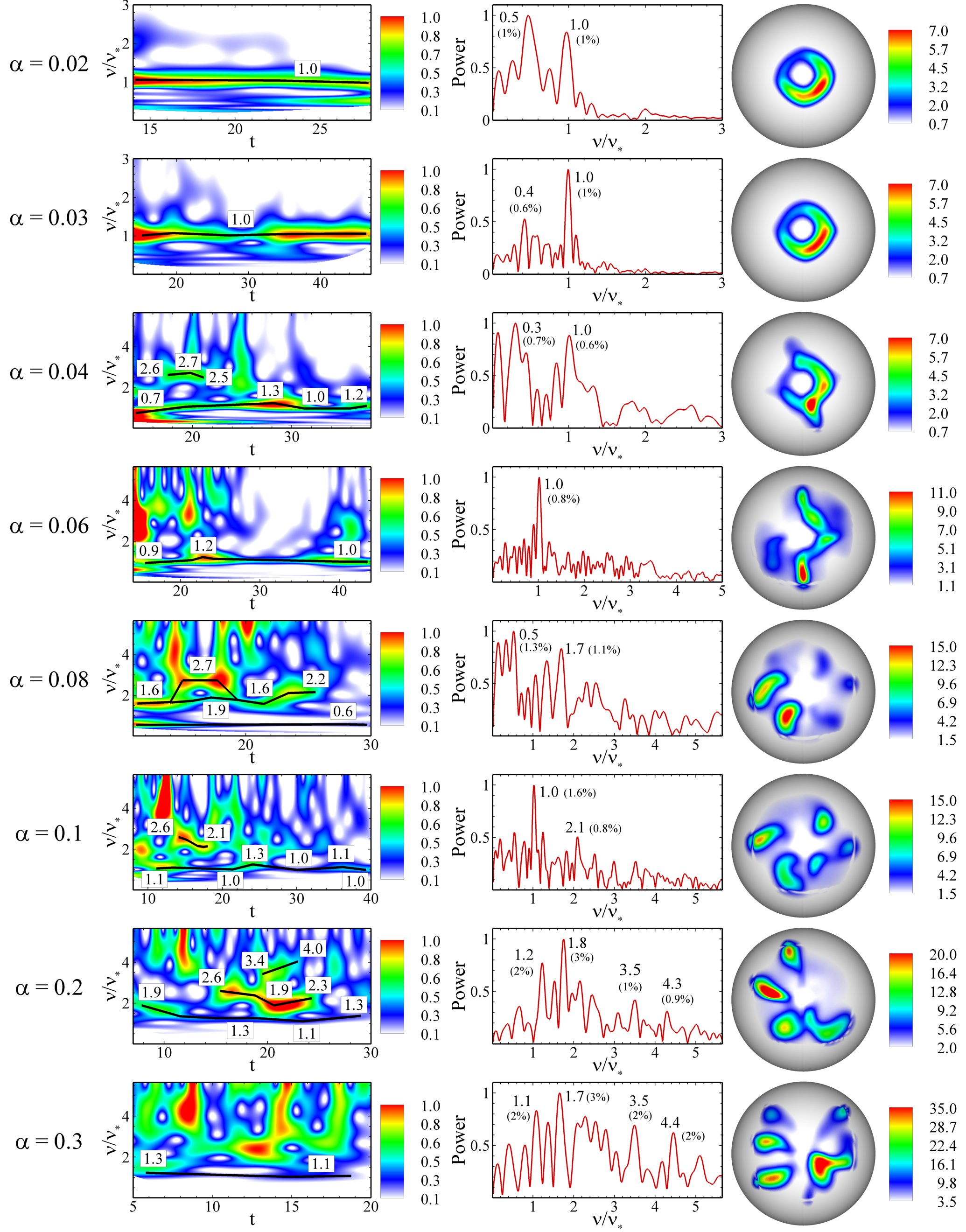}{wlet-alpha}{Wavelet and Fourier spectra and hotspots for cases that differ only in the accretion rate, controlled by the $\alpha$-viscosity parameter. In all these cases, $\Theta=5^\circ$, $P=2.8$ and $\mu=2$. The observer inclination angle is $i=45^\circ$. The numbers in the wavelet and Fourier plots are values of $\nu/\nu_*$. The colours in the hotspot and wavelet plots range from white (low) through blue to red (high).}

\subsection{Dependence of the Variability on the Accretion Rate}
\label{sec:wlet-alpha}
The accretion rate is one of the most important factors determining the accretion stability --- high accretion rates favour the instability \citep{KulkarniRomanova08}. This can be crudely understood as follows (for a more detailed discussion, see KR08 and, e.g., \citealt{SpruitStehlePapaloizou95} and \citealt{LiNarayan04} for instability criteria). The matter at the inner disk boundary is strongly braked by the stellar magnetic field and is brought closer to corotation with the star. The effective gravitational acceleration, which is the difference between the gravitational and centrifugal accelerations, is one of the important parameters which decide if the accretion is stable --- a stronger effective gravitational acceleration is favourable for the Rayleigh-Taylor instability. This acceleration is determined by how far the inner-disk angular velocity is from being Keplerian. When the accretion rate increases, the inner edge of the accretion disk moves inward, and the inner-disk matter becomes more strongly sub-Keplerian, increasing the effective gravitational acceleration and causing the instability to become stronger.

The most obvious way of controlling the accretion rate is through the disk density, but in our simulations, it is more convenient to vary the viscosity parameter $\alpha$ instead. We keep the dimensionless density fixed at $\rho_d=1$ in all our simulations. Increasing $\alpha$ increases the accretion rate, bringing the inner edge of the accretion disk inwards, as reflected by compression of the magnetosphere (\fig{xy-xz}, bottom row). However, increasing the density also makes the instability stronger (KR08, \S3.1). Thus, the reason the instability becomes stronger with increasing $\alpha$ is not the higher viscosity itself, but the higher accretion rate and the smaller resultant magnetospheric radius.

Therefore, to investigate the dependence of the instability and variability on the accretion rate, we vary $\alpha$, keeping the other parameters fixed at the following values: misalignment angle $\Theta=5^\circ$, stellar rotation period $P=2.8$ and stellar magnetic moment $\mu=2$. \fig{wlet-alpha} shows the wavelet and Fourier spectra for these cases. At $\alpha=0.02$ the accretion is stable, and the lightcurve has a peak at the star's rotation frequency (the peak at half that frequency is due to wandering of the hotspot itself, and is a separate issue). As we move on to $\alpha=0.03$, we start getting closer to the unstable regime. The wavelet spectrum shows the peak at the star's rotation frequency beginning to waver. This wavering is due to slight wandering of the funnel stream induced by the instability, although the instability is not strong enough to produce tongues at this stage. At $\alpha=0.04$ the accretion is unstable, and the peak near the star's rotation frequency is not steady; it is seen to drift between $\nu/\nu_*=0.7$ and $\nu/\nu_*=1.3$. The Fourier spectrum still shows a peak at $\nu/\nu_*=1.0$, but it is much broader and the spectrum has much more noise than in the stable accretion cases. Notice that the hotspot in this case has a banana-shaped portion produced by the funnels, and features protruding from the banana, which are created by tongues. The tongues, and hence these protrusions, rotate faster than the star, producing the high-frequency peaks at $\nu/\nu_*=2.5-2.7$ seen in the wavelet. There are two prominent protrusions, and they survive for a long time, rotating with an almost constant frequency. This is therefore a strong candidate for obtaining QPOs. We discuss this in more detail in \sect{spot-omega}.

As we move to higher accretion rates, we see that the noise in the Fourier spectra continues to increase, and hints of peaks at other frequencies start appearing. Note, however, that even though these cases are unstable, there is a reasonably clear peak at the star's rotation frequency, since accretion occurs through both funnels and tongues. It is only when we reach the strongly unstable case with $\alpha=0.2$ that the accretion occurs solely through tongues, and the star's rotation frequency disappears from the Fourier spectrum, which is dominated by other frequencies. At $\alpha=0.3$, it is difficult to discern any peaks in the Fourier spectrum. Observationally, this would correspond to lack of pulsations. We discuss the implications of this in \sect{conc}.

%---------------------------------------------------------

\subsection{More Sample Cases}
\label{sec:wlet-qpo}

\figwide{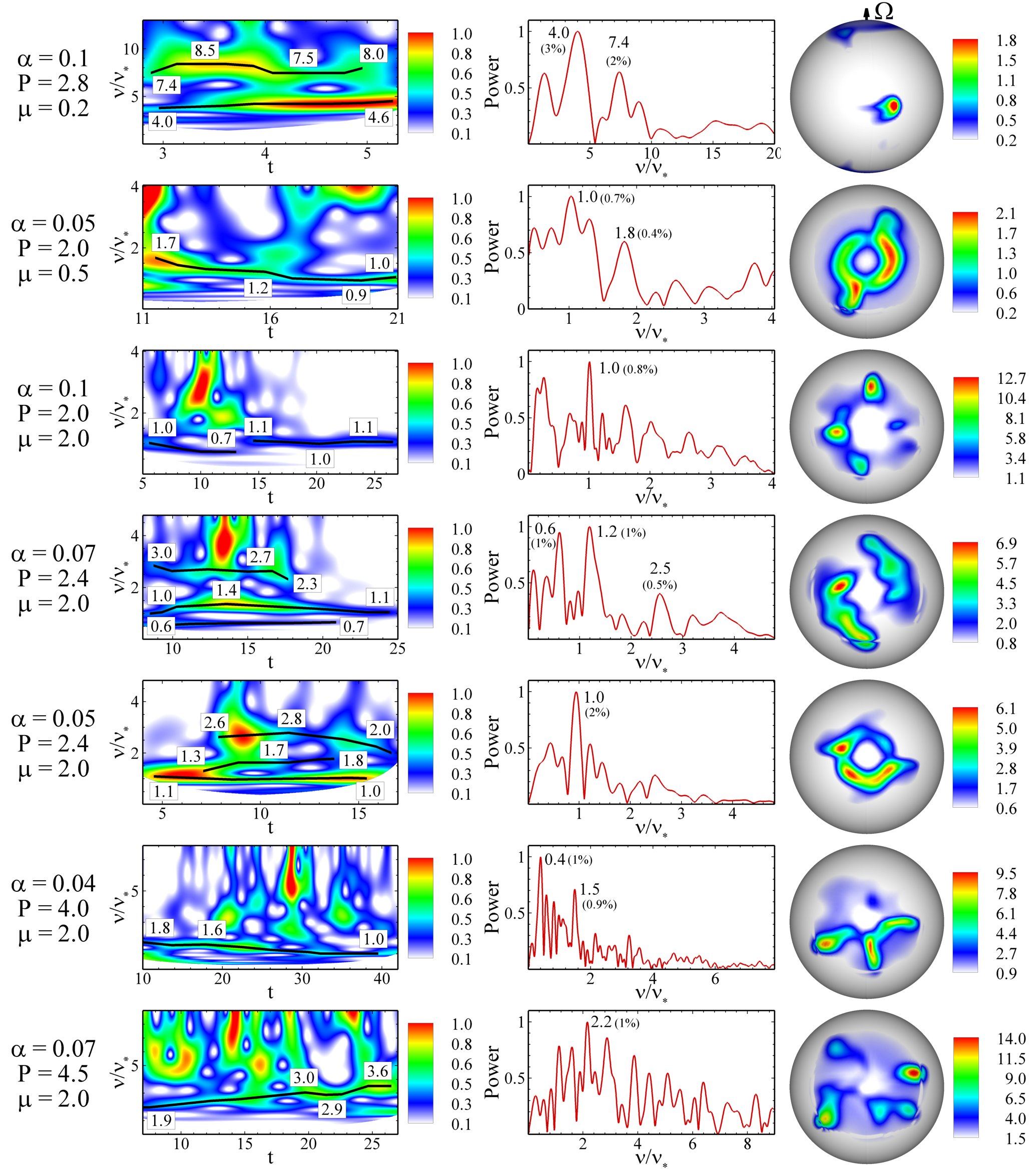}{wlet-qpo}{Wavelet and Fourier spectra and hotspots for various cases, showing QPO-like features. The observer inclination angle is $i=45^\circ$. The numbers in the wavelet and Fourier plots are values of $\nu/\nu_*$. The hotspots are all shown pole-on, except for the one in the top row, where the line of sight is in the plane of the rotational equator. The colours in the hotspot and wavelet plots range from white (low) through blue to red (high).}

\fig{wlet-qpo} shows some more cases with interesting spectral behaviour, for misalignment angle $\Theta=5^\circ$. The important thing to keep in mind here is that one of the important factors determing whether the instability exists is the difference between stellar rotation frequency and the Keplerian frequency at the inner disk radius, as mentioned in the previous section. Thus, increasing the rotation period is favourable for the instability.

The most interesting of the cases in \fig{wlet-qpo} is the one in the top row, which has the same parameters as in our main case (\fig{wlet-alpha}, $\alpha=0.1$), except for a very small magnetospheric radius ($r_m/R = 2$). The accretion is strongly unstable, occurring solely through tongues, and since the inner-disk orbital frequency is higher, the tongues rotate at a much higher frequency than in the main case. This produces peaks at very high frequencies in the power spectrum. The interesting thing about unstable cases with weak magnetic fields is that the behaviour of the tongues is noticeably different. There are usually two tongues, and they rotate almost rigidly with an almost constant angular velocity (\fig{lowb}), producing clear and steady QPOs. It must be pointed out, however, that the matter in the tongues does not rotate rigidly --- it is only the pattern of gas flow that does. We discuss accretion to small magnetospheres elsewhere \citep{RomanovaKulkarni08}.

The second row of \fig{wlet-qpo} shows another unstable case with two tongues that persist for a long time, rotating faster than the star with an almost constant angular velocity, like in the case with $\alpha=0.04$ described in \sect{wlet-alpha}. This is another strong candidate for obtaining QPOs.

The third row of \fig{wlet-qpo} shows an interesting phenomenon --- the instability exists only between $t=7$ and $t=14$. It is only during this period that the star's rotation frequency is absent from the wavelet spectrum. This may be relevant for the intermittency of pulsations from some accreting millisecond pulsars: changes in the accretion rate can cause the accretion to switch between being stable and unstable, causing the star's rotation frequency to alternately appear and disappear from the lightcurve power spectra. We discuss this in \S\ref{sec:conc}.

The wavelet plot in the fifth row shows an interesting feature: the existence of two relatively long-lived frequencies in addition to the stellar frequency. If they remain steady over longer periods of time, they can be expected to produce quasi-periodic oscillations.

The sixth and seventh rows show that even for relatively low values of the accretion rate, determined by $\alpha$, if the star rotates slowly, the instability can be strong enough that the star's rotation period is completely absent from the lightcurve power spectrum. 

\fignar{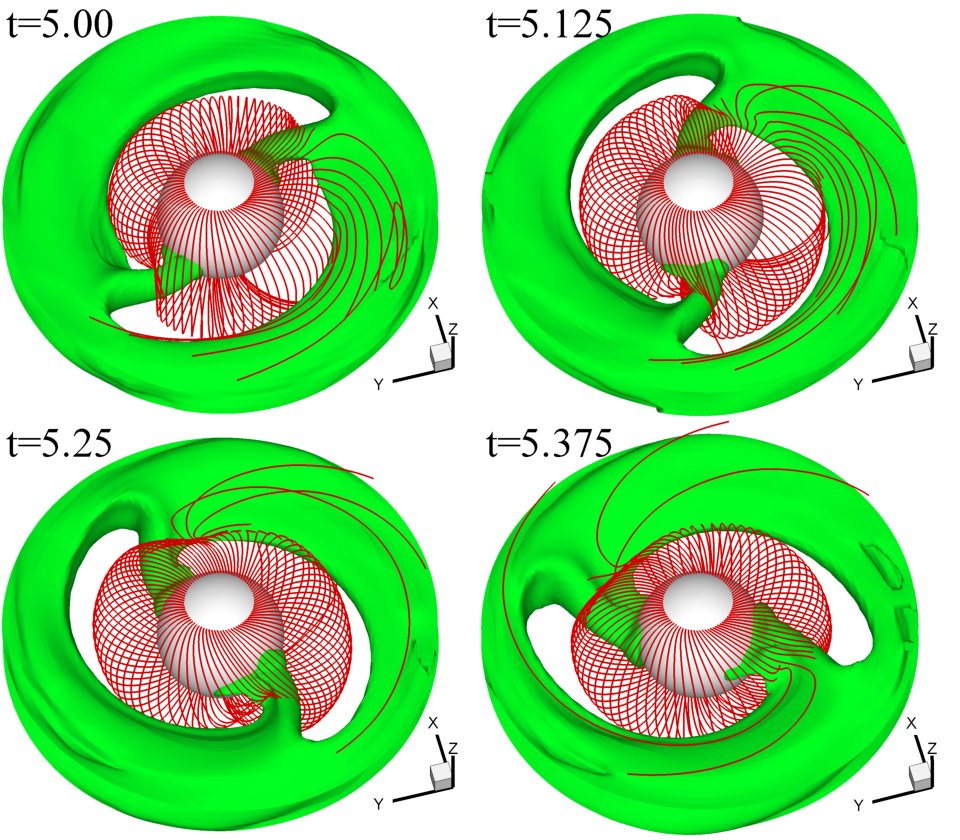}{lowb}{Motion of the tongues for a case with $\mu=0.2$, which has a small magnetosphere. Constant density surfaces and magnetic field lines are shown.}

% It must be emphasized, however, that we do not always see such frequency drifts. The cases shown in this and the preceding sections are meant only as examples of interesting behaviour in the frequency domain.

%---------------------------------------------------------

\subsection{Search for Explanation of Frequencies and Possible QPOs}
\label{sec:spot-omega}
We try to find an explanation for the various frequencies that appear during unstable accretion shown in the preceding sections, by tracking the motion of the hotspots on the star's surface. The hotspots produced by the tongues rotate approximately around the star's magnetic pole, so we choose a ring on the star's surface centered at the magnetic pole which contains the hotspots (\fig{spot-omega-method}). Then we integrate over the latitude $\theta$ the flux emitted from this ring, at each azimuthal position $\phi$ along the ring and each instant of time $t$, in a reference frame rotating with the star. This integrated flux is shown in \fig{spot-omega} for the cases with different values of the $\alpha$-viscosity shown in \fig{wlet-alpha}. The streaks marked by solid lines in this plot are produced by the hotspots, and the slopes of the streaks give the angular frequency of rotation of the spots. We therefore refer to these plots as ``spot-omega'' plots. The dashed lines indicate the continuation of the solid lines from $\phi=0$ when they reach $\phi=360^\circ$. The pair of numbers adjacent to each solid line denote the spot rotation frequency in units of the stellar rotation frequency (i.e., $\nu/\nu_*$), and the number of rotations performed by the spot during its lifetime, as seen by an external observer. The cases in which funnels are present show horizontal bands that simply indicate that the hotspots produced by the funnels stay at a fixed location on the star's surface, due to which they have $\nu/\nu_* = 1$ as seen by an external observer. These spots generally last the entire duration of the simulations; the number of rotations performed by them is therefore not shown. The other spots are produced by tongues, and rotate at approximately the orbital frequency of the inner-disk matter, which in turn depends on the Keplerian frequency at the inner disk radius. The angular frequencies of the spots may therefore be expected to change on long timescales if there are secular changes in the inner disk radius.

\fignar{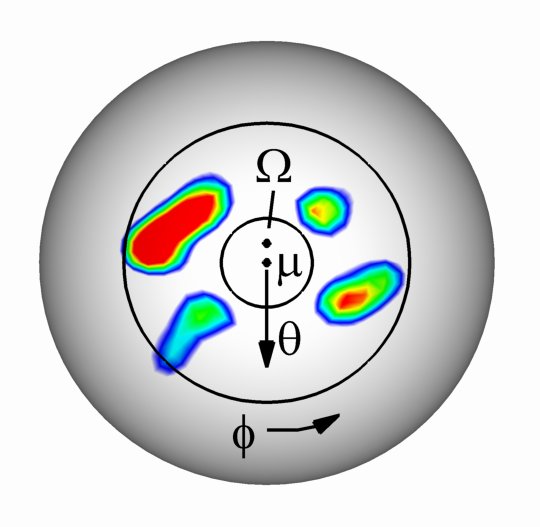}{spot-omega-method}{Setup for tracking the hotspot motion. The flux emitted from within the ring (bounded by the two circles and centered at the magnetic pole $\mu$) is integrated over the latitude $\theta$, at each longitude $\phi$ and each instant of time, giving the azimuthal position of each hotspot as a function of time. Zero longitude is defined as the direction away from the rotation axis $\Omega$.}

\figwide{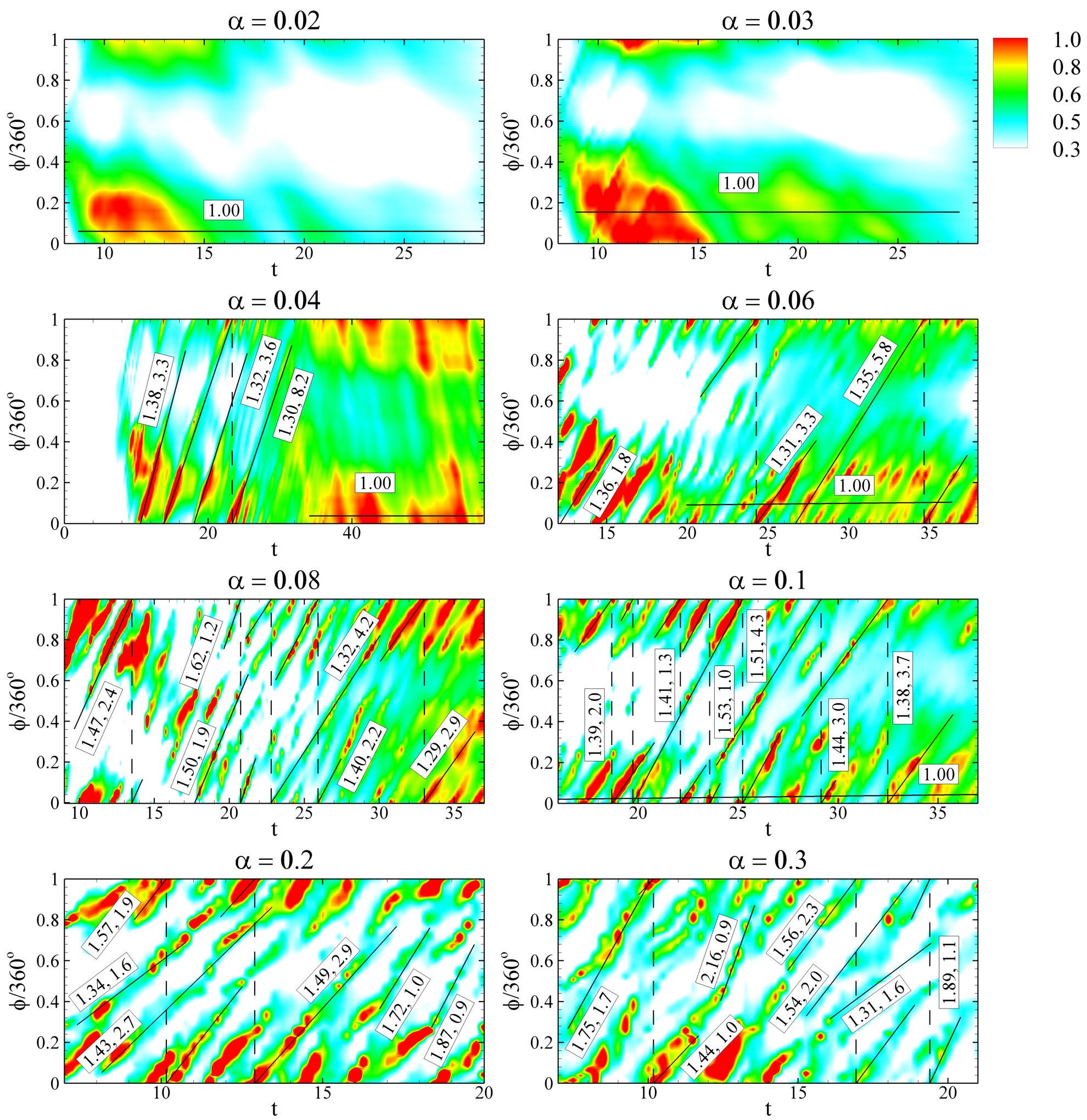}{spot-omega}{``Spot-omega'' plots to track hotspot motion. The normalized hotspot flux as a function of the magnetic longitude $\phi$ and time is shown for the cases with different $\alpha$ in \fig{wlet-alpha}, which have $\Theta=5^\circ$, $P=2.8$ and $\mu=2$. The first of each pair of numbers inside the plots is the value of $\nu/\nu_*$ for the hotspot, and the second is the number of rotations performed by the spot during its lifetime, as seen by an external observer. The colours range from white (low) through blue to red (high).}

There is some agreement between the frequencies obtained from the spot-omega plots and those from the fourier and wavelet plots. For example, the spot-omega plot for $\alpha=0.04$ shows $\nu/\nu_*\approx1.3$, which is close to some of the frequencies in the corresponding wavelet plot, and is also seen in the fourier plot, albeit not very distinctly. The spot-omega plot for the $\alpha=0.08$ case shows a range of frequencies between 1.2 and 1.6, which is consistent with the peaks seen in the fourier plot at that range of frequencies. The agreement, however, is not very encouraging. There are several possible reasons for this: (1) The hotspots produced by the tongues appear and disappear sporadically, and are relatively short-lived; as the spot-omega plots show, each spot only completes of the order of a few rotations around the star (as seen by an external observer). (2) Different spots separated either spatially or temporally are not perfectly correlated with regard to their location, again as shown by the spot-omega plots; the individual streaks in those plots show little correlation in location and duration. (3) The shape and brightness of the spots constantly changes. This is also seen from the spot-omega plots; the width and brightness of each streak varies as we move along the streak. (4) The angular velocity of the tongues, and therefore of the spots, is not constant; different tongues in the spot-omega plots have different angular velocities, and occasionally we even see a sudden jump in the frequency (e.g., for $\alpha=0.3$ at $t=12.5$). Due to these factors, the spots do not produce a coherent modulation of the observed flux. In fact, lack of coherent modulation, which consists of effects like rapid amplitude and phase changes, or ``jumps'' in the lightcurve, could produce short-duration peaks in the wavelet, which would appear in the Fourier spectra if the duration of the lightcurves is short, as in our simulations ($\sim$ 10 stellar rotation periods).

However, we believe that there is still a lot of hope of seeing the spot-omega frequencies in the lightcurves. The main culprit for the absence of these frequencies from the fourier and wavelet plots seems to be the short duration of our lightcurves combined with the lack of coherent modulation mentioned above. Since the tongues have a finite lifetime, their contribution to the lightcurve consists of wavetrains with random phase, amplitude and duration, and it is difficult to extract the frequencies of these wavetrains if the lightcurves are a few tens of stellar rotation periods long. However, the amplitude and duration distribution of these wavetrains is not very broad. Experiments with much longer, artificially generated lightcurves consisting of such wavetrains show clear presence of QPOs \citep{BachettiEtAl09}, due to the fact that in the fourier domain, the peaks due to definite signals grow faster than the noise when the length of the lightcurve increases. One encouraging fact about the spot-omega plots is that although the individual tongues are short-lived, the tongue frequency is relatively well-defined for a given set of parameters; the spread is generally not very large, except in the most strongly unstable cases ($\alpha\geq0.2$). This strongly increases the chances of obtaining QPOs from longer lightcurves. The spread in the hotspot frequency may contribute to the width of the QPO peak. The center of the peak would be determined by the center of this frequency distribution, which, as mentioned above, is very close to the inner disk frequency. We may therefore expect QPOs close to the inner disk frequency, except perhaps in the most strongly unstable cases.

Another interesting question concerns the significance of the frequencies in the fourier and wavelet plots. As mentioned earlier, a significant portion of them probably amount to nothing more than noise. However, it is possible that the hotspot shape and brightness changes have some quasi-periodicity that is difficult to detect using the spot-omega plots. The spot-omega plots do show some signs of this: the tongue spots are usually brighter around $\phi=0$, that is, when the hotspots, rotating around the magnetic pole, come close to the disk. This might produce some of the peaks in the wavelet and fourier plots. The other peaks would be expected to contribute to broadband noise in the power spectrum, the effect of which would a smaller relative amplitude of the peak at the star's rotation frequency than in the stable case. It is difficult to ascertain this without longer lightcurves.

It is also very interesting to observe the effect of increasing accretion rate (via increasing $\alpha$) on the tongue frequencies in the spot-omega plots. The most important effect is that the frequencies show an increasing trend. This is due to the fact that the disk comes in closer and the inner-disk orbital frequency, which determines the tongue frequency, increases. We also see signs of the tongue behaviour becoming more chaotic; the tongue lifetimes become shorter (although this is not a very clear trend), and the frequency spread increases. These two effects would be responsible for decreasing the strength of the QPO peak. Thus, is reasonable to expect that the most strongly unstable cases ($\alpha\geq0.2$) would not show any QPOs. The lightcurve power spectra in these cases would completely lack pulsations.

For stars with much smaller magnetospheres than those considered here, we find a different regime of accretion which we call the ``magnetic boundary layer regime,'' in which the behaviour of the tongues is much more coherent, producing well-defined peaks in the Fourier and wavelet spectra. We discuss this regime in detail in \citet{RomanovaKulkarni08}.

%---------------------------------------------------------

\subsection{Dependence of the Instability on the Misalignment Angle}
\label{sec:theta-dep}

\fignar{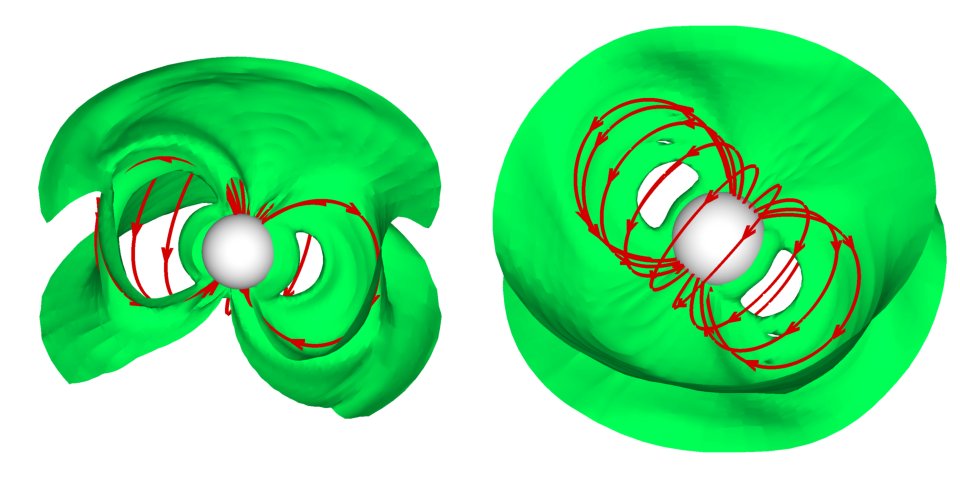}{funnel-tongue}{{\it Left panel:} Cutaway view of the accretion flow in an unstable case with $\Theta=30^\circ$. {\it Right panel:} Accretion flow in an unstable case with $\Theta=60^\circ$.}

\figwide{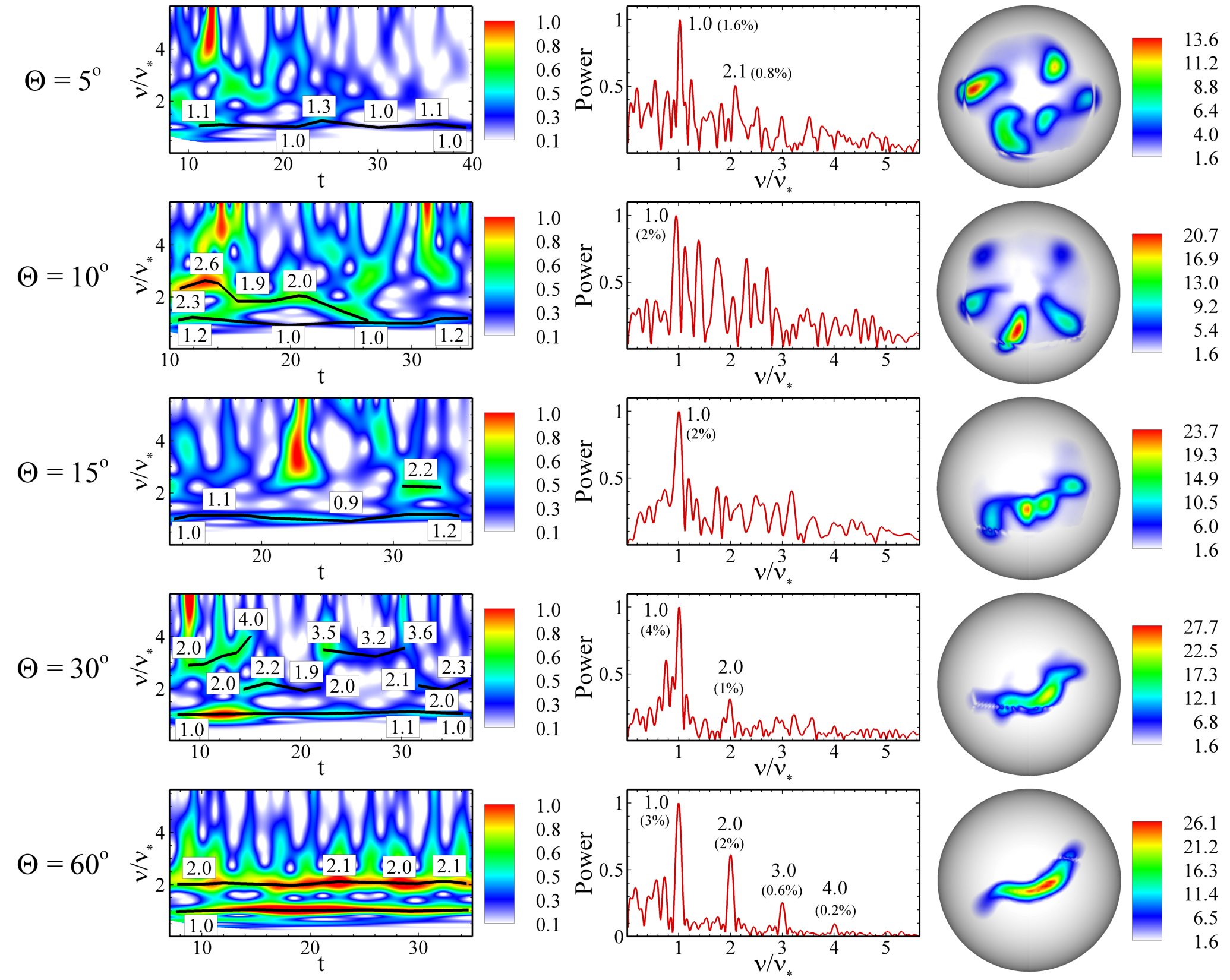}{wlet-theta}{Wavelet and Fourier spectra and hotspots for cases that differ only in the misalignment angle $\Theta$. In all these cases, $\alpha=0.1$, $P=2.8$ and $\mu=2$. The observer inclination angle is $i=45^\circ$. The numbers in the wavelet and Fourier plots are values of $\nu/\nu_*$. The colours in the hotspot and wavelet plots range from white (low) through blue to red (high).}

So far we have considered cases with a small misalignment angle, $\Theta=5^\circ$. For higher $\Theta$ the accretion geometry is slightly different even in the stable accretion regime, due to the stronger non-axisymmetry. The funnel flows are closer to the equatorial plane and produce elongated, rather than banana-shaped, hotspots (\citealt{RomanovaEtAl04}; KR05). It is therefore interesting to explore the effect of $\Theta$ on the unstable accretion geometry. At high misalignment angles ($\Theta\gtrsim25^\circ$), the funnels are always present. In this situation, in the unstable cases, the tongues have low density, and are restricted to being under the edges of the funnels, due to which the funnels appear to be ``connected'', as \fig{funnel-tongue} shows. This connection appears and disappears stochastically. As we attempt to move into the strongly unstable regime (by either increasing the accretion rate or decreasing the star's rotation rate), the connecting tongues increase in density and become permanent. The magnetic field lines near the edges of the funnels are pried apart by these tongues and stay in that configuration, creating a new equilibrium configuration with a stable accretion flow, albeit partially inside the magnetosphere.

To explore the effect of $\Theta$ on the variability properties, we first vary $\Theta$ keeping the other parameters fixed at the following values: viscosity parameter $\alpha=0.1$, stellar rotation period $P=2.8$ and stellar magnetic moment $\mu=2$. \fig{wlet-theta} shows the effect on the power spectrum of increasing $\Theta$, which is similar to decreasing the accretion rate. For the parameters chosen here, funnels are present for all values of $\Theta$. For small $\Theta$, the peak at the star's rotation frequency is visible, but the power spectrum is noisy due to the presence of the tongues. As we go to higher $\Theta$, the motion of the instability tongues is azimuthally restricted as mentioned above. The hotspots are therefore more or less fixed on the star's surface, and so the stellar rotation frequency becomes clearer in the lightcurve.

Notice the high harmonic content of the power spectrum for $\Theta=60^\circ$. This is probably due to the fact since the misalignment angle is large, the hotspot (which is close to the magnetic pole) is close to the rotational equator, and therefore has a higher linear velocity, as seen by an external observer, than for smaller misalignment angles. The distortion of the lightcurve away from a sinusoidal shape due to the Doppler effect is therefore stronger (see, e.g., \citealt{PoutanenGierlinski03}; KR05).

From the point of view of observations, it is interesting to know whether or not the lightcurves show periodicity at the star's rotation frequency. In order to examine in detail the parameter ranges for which they do, we performed a range of simulation runs for different values of the misalignment angle $\Theta$ and the stellar rotation period $P$. We chose the stellar magnetic moment to be $\mu=1$ and the viscosity parameter to be $\alpha=0.02$ in all these runs. \fig{theta-dep-a0.02} shows the results. The lightcurve lacks periodicity at the star's rotation period only below a critical misalignment angle $\Theta_{crit}$ which depends on the rotation rate. We see that for $P\lesssim 2.2$, $\Theta_{crit}<0$; the star's period is visible for all misalignment angles. As we decrease the star's rotation rate, the difference between the star's rotation rate and the Keplerian frequency at the inner disk radius increases, causing stronger magnetic braking of the matter. Hence, the effective gravitational acceleration at the inner disk edge increases, causing the instability to become stronger, and $\Theta_{crit}$ to increase. However, when $\Theta_{crit}$ reaches $\sim 25^\circ$, it stops rising. For larger misalignment angles than this, the funnels are always present, and their contribution to the hotspots gives rise to variability at the star's rotation frequency.

\fignar{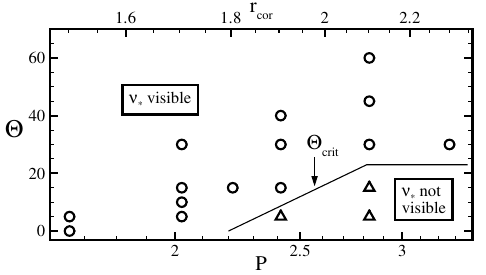}{theta-dep-a0.02}{The regimes in which the lightcurve shows the star's rotation period, as a function of the misalignment angle $\Theta$ (in degrees) and rotation period $P$, for $\alpha=0.02$ and $\mu=1$. The circles and triangles represent simulation runs that have lightcurves with and without the star's rotation period respectively.}

%---------------------------------------------------------

\subsection{Dependence of Results on Model Parameters}
One question that remains concerns the dependence of the results described above on the viscosity parameter $\alpha$ and the grid resolution. The effect of these parameters on the behaviour of the tongues has been discused in detail in KR08. To summarize, one of the important factors that determine the appearance of the instability is the balance between the gravitational and centrifugal forces at the inner disk boundary, which in turn depends on the accretion rate and the stellar rotation rate and magnetic field. The viscosity parameter comes into play only insofar as it determines the accretion rate, and does not impact the variability properties directly. To check this, we chose a case with a small value of $\alpha$ where the accretion was stable, and increased the density in the disk, to find that the accretion became unstable (KR08, \S3.1). Similarly, from the last two rows of \fig{wlet-qpo} we see that in spite of the relatively low values of $\alpha$, the instability exists and is so strong that there is no accretion through funnels, as evidenced by the absence of the star's rotation frequency from the power spectra. Thus, it is the accretion rate, and not the viscosity, which is important for determining the properties of the instability. The viscosity coefficient does, however, impact the radial velocity of the accreting matter; the effect, if any, of this on the instability needs to be explored. As far as the dependence on the grid resolution is concerned, the important thing to note is that the aspects of the tongue dynamics which determine the variability properties are the number and rotation rate of the tongues, and to a lesser degree, the location on the star where the tongues deposit matter. These properties are found to be independent of the grid resolution (KR08, \S3.4).
%=========================================================

\section{Conclusions and Discussion}
\label{sec:conc}

The main results of this work are the following:

\begin{enumerate}
\item For small misalignment angles between the rotation and magnetic axes of the star, the tongues that penetrate the magnetosphere during unstable accretion, and the resulting hotspots on the star's surface, have an orbital frequency close to that of the inner-disk in most cases. We therefore expect to obtain quasi-periodic oscillations (QPOs) at the inner-disk orbital frequency in a wide range of unstably accreting systems. As we move deeper into the unstable regime, the tongue behaviour becomes more and more chaotic, and we may expect the QPOs to become weaker and gradually disappear.

\item There is significant accretion through antipodal funnels even during unstable accretion for a broad range of parameter values. The lightcurves therefore show periodicity at the stellar rotation frequency or twice that, depending on the misalignment angle and the viewing geometry, as in cases with stable accretion.

\item For large misalignment angles, the accretion funnels are always present, and the tongue motion is restricted, due to which the power spectra always show the star's rotation frequency.

\end{enumerate}

\begin{itemize}
\item {\bf\boldmath At small misalignment angles ($\Theta\lesssim25^\circ$)} and magnetospheric sizes of a few stellar radii, the hotspots due to the tongues exhibit stochastic behaviour, and thus contribute only noise to the power spectra on the timescales explored here. The important thing, however, is that the rotation frequency of the tongues and the resulting hotspots stays relatively steady at a value close to the inner-disk orbital frequency, even in the nonlinear regime of the instability, except in the most strongly unstable cases. This could produce QPOs in the power spectra at frequencies close to the inner-disk orbital frequency. Longer lightcurves are needed to verify this. The stochastic behaviour of the tongues may contribute to low-frequency broadband noise in the power spectra \citep[see, e.g.,][]{BurderiEtAl97}. We see some signs of this in our power spectra; the noise amplitude is lower at high frequencies in most cases. At the same time, due to the presence of the funnels, the star's frequency is expected to appear in the power spectra in most unstable cases.

There exists a strongly unstable regime of accretion, however, in which the funnels are absent, and the lightcurves do not show periodicity at the star's rotation frequency; moreover, the tongue behaviour is more chaotic, making the presence of QPOs much less likely. These cases may correspond to accreting neutron stars without any pulsations. Since unstable accretion occurs at relatively high accretion rates (\sect{wlet-alpha}; also KR08), the above results have a few implications: (1) If the accretion rate is close to the boundary between stable and unstable regimes, slight changes in the accretion rate can cause the accretion to episodically switch between being stable and unstable, causing corresponding appearance and disappearance of pulsations (\sect{wlet-qpo}). This might be a possible explanation for the behaviour of intermittent pulsars \citep[e.g.,][]{AltamiranoEtAl08, CasellaEtAl08}. One of the attractive features of this idea comes from the fact that during stable accretion, the azimuthal location of funnels with respect to the star is fixed, even across periods of unstable accretion. The pulsations in intermittent pulsars would therefore be expected to be coherent in phase across periods of lack of pulsations. This ``phase memory'' has been observed in the intermittent pulsars mentioned above. (2) Lack of pulsations does not imply a dynamically important magnetic field. The most strongly unstable cases shown in this work lack pulsations in their lightcurves, but have distinct magnetospheres that strongly affect the matter flow around the star. The magnetic moment in these cases is stronger by a factor of $\sim 20$ than in cases that show field burial, discussed in \citet{RomanovaKulkarni08}. Field burial is an argument that is often advanced as an explanation for the paucity of accreting millisecond pulsars. The unstable regime complements this argument by providing another way for pulsations to be absent.

\item {\bf\boldmath At large misalignment angles ($\Theta\gtrsim25^\circ$)}, funnels are always present, and the motion of the tongues is azimuthally restricted to being close to the edges of the funnels. The hotspots are therefore approximately fixed on the star's surface, due to which the lightcurves always show the stellar rotation frequency, even during strongly unstable accretion.

\end{itemize}

Although this work focuses on neutron stars, we expect the above conclusions to be applicable to other types of accreting magnetized stars in a broad sense. One thing needs to be noted, however: the magnetospheres in all the simulations presented here are a few stellar radii in size. For much smaller magnetospheres, the variability is very different \citep{RomanovaKulkarni08}. In the stable regime, the funnels show significant wandering, due to which the star's rotation frequency is absent from the lightcurve power spectra. In the unstable regime, the tongues are much more coherent, producing distinct peaks in the power spectra.

%#########################################################

\subsection*{Acknowledgments} The authors wish to thank Matteo Bachetti for valuable discussions, and the referee for various comments that improved the presentation of this paper. The authors were supported in part by NASA grant NNX08AH25G and by NSF grants AST-0607135 and AST-0807129. We are thankful to NASA for access to the NASA High Performance Computing Facilities.

%#########################################################

\bibliography{ms}

\begin{thebibliography}{}

\bibitem[\protect\citeauthoryear{{Altamirano}, {Casella}, {Patruno}, {Wijnands}
  \& {van der Klis}}{{Altamirano} et~al.}{2008}]{AltamiranoEtAl08}
{Altamirano} D.,  {Casella} P.,  {Patruno} A.,  {Wijnands} R.,    {van der
  Klis} M.,  2008, \apjl, 674, L45

\bibitem[\protect\citeauthoryear{{Arons} \& {Lea}}{{Arons} \&
  {Lea}}{1976}]{AronsLea76}
{Arons} J.,  {Lea} S.~M.,  1976, \apj, 207, 914

\bibitem[\protect\citeauthoryear{{Bachetti} \& {et al.}}{{Bachetti} \& {et
  al.}}{2009}]{BachettiEtAl09}
{Bachetti} M.,  {et al.} 2009, in preparation

\bibitem[\protect\citeauthoryear{{Beloborodov}}{{Beloborodov}}{2002}]{Beloboro%
dov02}
{Beloborodov} A.~M.,  2002, \apjl, 566, L85

\bibitem[\protect\citeauthoryear{{Bouvier}, {Alencar}, {Harries}, {Johns-Krull}
  \& {Romanova}}{{Bouvier} et~al.}{2007}]{BouvierEtAl07}
{Bouvier} J.,  {Alencar} S.~H.~P.,  {Harries} T.~J.,  {Johns-Krull} C.~M.,
  {Romanova} M.~M.,  2007, in {Reipurth} B.,  {Jewitt} D.,   {Keil} K.,  eds,
  Protostars and Planets V {Magnetospheric Accretion in Classical T Tauri
  Stars}.
pp 479--494

\bibitem[\protect\citeauthoryear{{Burderi}, {Robba}, {La Barbera} \&
  {Guainazzi}}{{Burderi} et~al.}{1997}]{BurderiEtAl97}
{Burderi} L.,  {Robba} N.~R.,  {La Barbera} N.,    {Guainazzi} M.,  1997, \apj,
  481, 943

\bibitem[\protect\citeauthoryear{{Casella}, {Altamirano}, {Patruno}, {Wijnands}
  \& {van der Klis}}{{Casella} et~al.}{2008}]{CasellaEtAl08}
{Casella} P.,  {Altamirano} D.,  {Patruno} A.,  {Wijnands} R.,    {van der
  Klis} M.,  2008, \apjl, 674, L41

\bibitem[\protect\citeauthoryear{{Donati}, {Jardine}, {Gregory}, {Petit},
  {Bouvier}, {Dougados}, {M{\'e}nard}, {Cameron}, {Harries}, {Jeffers} \&
  {Paletou}}{{Donati} et~al.}{2007}]{DonatiEtAl07}
{Donati} J.-F.,  {Jardine} M.~M.,  {Gregory} S.~G.,  {Petit} P.,  {Bouvier} J.,
   {Dougados} C.,  {M{\'e}nard} F.,  {Cameron} A.~C.,  {Harries} T.~J.,
  {Jeffers} S.~V.,    {Paletou} F.,  2007, \mnras, 380, 1297

\bibitem[\protect\citeauthoryear{{Elsner} \& {Lamb}}{{Elsner} \&
  {Lamb}}{1977}]{ElsnerLamb77}
{Elsner} R.~F.,  {Lamb} F.~K.,  1977, \apj, 215, 897

\bibitem[\protect\citeauthoryear{{Ghosh} \& {Lamb}}{{Ghosh} \&
  {Lamb}}{1979}]{GhoshLamb79}
{Ghosh} P.,  {Lamb} F.~K.,  1979, \apj, 232, 259

\bibitem[\protect\citeauthoryear{{Koldoba}, {Romanova}, {Ustyugova} \&
  {Lovelace}}{{Koldoba} et~al.}{2002}]{KoldobaEtAl02}
{Koldoba} A.~V.,  {Romanova} M.~M.,  {Ustyugova} G.~V.,    {Lovelace} R.~V.~E.,
   2002, \apjl, 576, L53

\bibitem[\protect\citeauthoryear{{K\"onigl}}{{K\"onigl}}{1991}]{Konigl91}
{K\"onigl} A.,  1991, \apjl, 370, L39

\bibitem[\protect\citeauthoryear{{Kulkarni} \& {Romanova}}{{Kulkarni} \&
  {Romanova}}{2005}]{KulkarniRomanova05}
{Kulkarni} A.~K.,  {Romanova} M.~M.,  2005, \apj, 633, 349

\bibitem[\protect\citeauthoryear{{Kulkarni} \& {Romanova}}{{Kulkarni} \&
  {Romanova}}{2008}]{KulkarniRomanova08}
{Kulkarni} A.~K.,  {Romanova} M.~M.,  2008, \mnras, 386, 673

\bibitem[\protect\citeauthoryear{{Li} \& {Narayan}}{{Li} \&
  {Narayan}}{2004}]{LiNarayan04}
{Li} L.-X.,  {Narayan} R.,  2004, \apj, 601, 414

\bibitem[\protect\citeauthoryear{{Poutanen} \& {Gierli{\'n}ski}}{{Poutanen} \&
  {Gierli{\'n}ski}}{2003}]{PoutanenGierlinski03}
{Poutanen} J.,  {Gierli{\'n}ski} M.,  2003, \mnras, 343, 1301

\bibitem[\protect\citeauthoryear{{Romanova} \& {Kulkarni}}{{Romanova} \&
  {Kulkarni}}{2008}]{RomanovaKulkarni08}
{Romanova} M.~M.,  {Kulkarni} A.~K.,  in preparation, 2008

\bibitem[\protect\citeauthoryear{{Romanova}, {Kulkarni} \&
  {Lovelace}}{{Romanova} et~al.}{2008}]{RomanovaKulkarniLovelace08}
{Romanova} M.~M.,  {Kulkarni} A.~K.,    {Lovelace} R.~V.~E.,  2008, \apjl, 673,
  L171

\bibitem[\protect\citeauthoryear{{Romanova}, {Ustyugova}, {Koldoba} \&
  {Lovelace}}{{Romanova} et~al.}{2004}]{RomanovaEtAl04}
{Romanova} M.~M.,  {Ustyugova} G.~V.,  {Koldoba} A.~V.,    {Lovelace} R.~V.~E.,
   2004, \apj, 610, 920

\bibitem[\protect\citeauthoryear{{Romanova}, {Ustyugova}, {Koldoba}, {Wick} \&
  {Lovelace}}{{Romanova} et~al.}{2003}]{RomanovaEtAl03}
{Romanova} M.~M.,  {Ustyugova} G.~V.,  {Koldoba} A.~V.,  {Wick} J.~V.,
  {Lovelace} R.~V.~E.,  2003, \apj, 595, 1009

\bibitem[\protect\citeauthoryear{{Spruit}, {Stehle} \& {Papaloizou}}{{Spruit}
  et~al.}{1995}]{SpruitStehlePapaloizou95}
{Spruit} H.~C.,  {Stehle} R.,    {Papaloizou} J.~C.~B.,  1995, \mnras, 275,
  1223

\bibitem[\protect\citeauthoryear{{van der Klis}}{{van der
  Klis}}{2004}]{vanderKlis04}
{van der Klis} M.,  2004, ArXiv Astrophysics e-prints

\bibitem[\protect\citeauthoryear{{Wang} \& {Robertson}}{{Wang} \&
  {Robertson}}{1985}]{WangRobertson85}
{Wang} Y.-M.,  {Robertson} J.~A.,  1985, \apj, 299, 85

\bibitem[\protect\citeauthoryear{{Warner}}{{Warner}}{1995}]{Warner95}
{Warner} B.,  1995, {Cataclysmic variable stars}.
Cambridge Astrophysics Series, Cambridge, New York: Cambridge University Press,
  |c1995

\bibitem[\protect\citeauthoryear{{Warner}, {Woudt} \& {Pretorius}}{{Warner}
  et~al.}{2003}]{WarnerWoudtPretorius03}
{Warner} B.,  {Woudt} P.~A.,    {Pretorius} M.~L.,  2003, \mnras, 344, 1193

\end{thebibliography}

\appendix

\section{Reference Values}
\label{app:refval}

\begin{table}
\begin{tabular}{l@{\extracolsep{0.2em}}l@{}lll}

\hline
&                                                   & CTTSs       & White dwarfs          & Neutron stars           \\
\hline

\multicolumn{2}{l}{$M(M_\odot)$}                    & 0.8         & 1                     & 1.4                     \\
\multicolumn{2}{l}{$R$}                             & $2R_\odot$  & 5000 km               & 10 km                   \\
\multicolumn{2}{l}{$R_0$ (cm)}                      & $4\e{11}$   & $1.4\e9$              & $2.9\e6$                \\
\multicolumn{2}{l}{$v_0$ (cm s$^{-1}$)}             & $1.6\e7$    & $3\e8$                & $8.1\e9$                \\
\multicolumn{2}{l}{$\omega_0$ (s$^{-1}$)}           & $4\e{-5}$   & 0.2                   & $2.8\e3$                \\
\multicolumn{2}{l}{\multirow{2}{*}{$P_0$}}          & $1.5\e5$ s  & \multirow{2}{*}{29 s} & \multirow{2}{*}{2.2 ms} \\
&                                                   & $=1.8$ days &                       &                         \\
\multicolumn{2}{l}{$B_{\star_0}$ (G)}               & $10^3$      & $10^6$                & $10^9$                  \\
\multicolumn{2}{l}{$B_0$ (G)}                       & 43          & $4.3\e4$              & $4.3\e7$                \\
\multicolumn{2}{l}{$\rho_0$ (g cm$^{-3}$)}          & $7\e{-12}$  & $2\e{-8}$             & $2.8\e{-5}$             \\
\multicolumn{2}{l}{$p_0$ (dy cm$^{-2}$)}            & $1.8\e{3}$  & $1.8\e{9}$            & $1.8\e{15}$             \\
\multirow{2}{*}{$\dot M_0$} & (g s$^{-1}$)          & $1.8\e{19}$ & $1.2\e{19}$           & $1.9\e{18}$             \\
                            & ($M_\odot$yr$^{-1}$)  & $2.8\e{-7}$ & $1.9\e{-7}$           & $2.9\e{-8}$             \\
\multicolumn{2}{l}{$T_0$ (K)}                       & $1.6\e6$    & $5.6\e8$              & $3.9\e{11}$             \\
\multicolumn{2}{l}{$\dot E_0$ (erg s$^{-1}$)}       & $4.8\e{33}$ & $1.2\e{36}$           & $1.2\e{38}$             \\
\multicolumn{2}{l}{$T_{\mathrm{eff},0}$ (K)}        & 4800        & $3.2\e5$              & $2.3\e7$                \\
\hline
\end{tabular}
\caption{Sample reference values of the dynamical quantities used in our simulations.}
\label{tab:refval}
\end{table}

As stated in section \sect{refval}, our simulations are done using dimensionless variables, obtained by dividing the dimensional variables by their respective reference values. The reference values are determined as follows: The unit of distance $R_0$ is chosen such that the star has radius $R = 0.35R_0$. The reference velocity is the Keplerian velocity at $R_0$, $v_0 = (GM/R_0)^{1/2}$, and $\omega_0 = v_0/R_0$ is the reference angular velocity.
The reference time is the Keplerian rotation period at $R_0$, $P_0 = 2\pi R_0/v_0$. The reference surface magnetic field of the star at the magnetic equator is $B_{\star_0}$. The reference magnetic field, $B_0$, is the initial magnetic field strength at $r=R_0$, assuming a surface magnetic field of $B_{\star_0}$. The reference magnetic dipole moment is $\mu_0 = B_{*0} R_*^3 \equiv B_0 R_0^3$. A dimensionless magnetic moment of $\mu'$ then corresponds to a surface magnetic field of $B_* = \mu'B_{*0}$. The reference density is taken to be $\rho_0 = B_0^2/v_0^2$. The reference pressure is $p_0 = \rho_0 v_0^2$. The reference temperature is $T_0 = p_0/{\mathcal R}\rho_0$, where $\mathcal{R}$ is the gas constant. The reference accretion rate is $\dot{M}_0 = \rho_0 v_0 R_0^2$. The reference energy flux is $\dot{E}_0 = \rho_0 v_0^3R_0^2$. The reference value for the effective blackbody temperature of the hot spots is $(T_{\mathrm eff})_0 = (\rho_0 v_0^3/\sigma)^{1/4}$, where $\sigma$ is the Stefan-Boltzmann constant. Table \ref{tab:refval} shows sample reference values for three classes of objects: classical T Tauri stars (CTTSs), white dwarfs and neutron stars.

\end{document}